\documentclass[twocolumn,final]{svjour3}

\newcommand{\ddt}[1]{\frac{\mathrm{d}#1}{\mathrm{d}t}}

\usepackage[latin1]{inputenc}
\usepackage{graphicx,graphics,amsmath,amssymb,stmaryrd,bm,dcolumn,textcomp,braket}
\usepackage{mathrsfs}
\usepackage{rotating,array,color,float}
\usepackage{xcolor,siunitx}
\usepackage{hyperref}
\usepackage[inline]{showlabels}
\hypersetup{colorlinks,linkcolor={blue!80!black},citecolor={blue!80!black},urlcolor={blue!80!black}}

\title{On the methodologies for the assessment of the impact of parameters in acoustophoretic separation devices}

\author{Fabio Garofalo}
\institute{Department of Biomedical Engineering, Lund University\\
Ole R\"omers V\"ag 3 S-22363, Lund, Sweden\\
\email{fabio.garofalo@bme.lth.se}
}
\authorrunning{Garofalo F.}
\titlerunning{a}

\begin{document}

\maketitle

\begin{abstract}
In this communication I reconcile the kinematic method illustrated by some authors~\cite{yang2018,vitali2018} in studying the impact of system and suspension parameters on acoustophoretic separations with the statistical method formerly proposed by Garofalo~\cite{garofalo2014,garofalo2014_2} and lately extended to particle populations by the same author~\cite{garofalo2017,garofalo2018}.
The connection between these two methods is established by (i)~reinterpreting the kinematic method in terms of tangent space dynamics, and (ii)~transforming the dynamics in the tangent space into the dynamics of the area elements.
The dynamics of the area elements is equivalent to the dynamics of the covariance matrix derived by moment analysis and associated with the dispersion problem during microparticle acoustophoresis.
The similarities and the differences between the kinematic based method and the stochastic method proposed by the present auuthor are illustrated and discussed in the light of the numerical results for a prototypical model of acoustophoretic separation.
\end{abstract}

\section{Introduction}
Some authors~\cite{yang2018,vitali2018} have investigated the influence of the particle and system parameters on the acoustophoretic separation of microparticles by considering the integration of the equation of motion for a particle suspended in a medium
\begin{equation}\label{eq:motion}
\ddt{\bm{x}}=\bm{v}(\bm{x},\bm{p})\,,\qquad \bm{x}(0)=\bm{x}_0^{}\,,
\end{equation}
where \mbox{$\bm{x}=[x,y,z]^T_{}$} is the particle position in the three-dimensional space, \mbox{$\bm{x}_0^{}=[x_0^{},y_0^{},z_0^{}]^T_{}$} the initial position, and \mbox{$\bm{p}=[p_1^{},...,p_N^{}]^T_{}$} is a set of $N$ parameters.

This set includes all the parameters that matter in quantifying the acoustophoretic separation: particle properties, fluid properties, and device properties. Yang et al. subdivided the parameters in intrinsic and extrinsic, this subdivision is not used here.

Since in~\cite{yang2018,vitali2018} the authors aimed to develop the analysis for quantifying the separation of biological samples, they suggested, by using the words ``dispersed objects'', ``populations'' or ``average position'', that the statistics of the sample was taken into account to some extent.
However, in their paper they did not consider any statistical analysis of dispersion during the dynamics of the separation process.
Rather, they considered two (or more) extremal values for the initial positions or parameters and studied the deviation of two trajectories corresponding to the ``worst'' and ``best'' cases. This led to a troublesome analysis (see Eqs.(9) in~\cite{vitali2018}) where the number of equations is quite large. However, the problem of studying the influence of small deviations from initial conditions is well-known in dynamic system theory.
Furthermore, founding their approach on this kind of methodology, they heuristically defined indicators to quantify the separation, such as the ``relative displacement'', the ``ideal separation efficiency'', and the ``separation efficiency''.
These indicators resulted labored reshapings of distances along the separation direction, i.e. the $y$-direction, between the kinematic trajectories resulting from the integration of equation of motion for the extremal values of the particle/system parameters.

Furthermore, it must be noticed that in citing Garofalo~\cite{garofalo2014} and by addressing that ``the theoretical analysis of particle kinematics in free-flow acoustophoretic devices was already reported'', they definitively overlooked the opportunity to discuss the results of their study of dispersion against a seemingly different method for the study of the ``cross-sectional dispersion''.
Indeed, it is noticeable the similarity between the behavior of the bandwidth~\cite{yang2018,vitali2018} and that of the variance formerly presented in~\cite{garofalo2014} and successively in~\cite{garofalo2017,garofalo2018}.
This affinity goes beyond the bandwidth and the variance when one compares the graphs of the trajectories/bandwidth in~\cite{yang2018} and those for the spatial distributions in~\cite{garofalo2014}.
This link went unnoticed and missing in the discussion of the results presented by the authors in~\cite{yang2018,vitali2018}.

This communication aims to compare and reconcile the approach in~\cite{yang2018,vitali2018} with the statistic-based method for studying dispersion during acoustophoretic separation proposed by Garofalo~\cite{garofalo2014,garofalo2014_2,garofalo2017,garofalo2018}.

\section{Theoretical Comparison}\label{sec:theory}
\subsection{Dynamics}\label{sec:similarities}
Let us consider the extension of Eq.~\eqref{eq:motion} in the coordinates+parameters space by including a dummy dynamics in the parameters, i.e. \mbox{$\bm{p}(t)=\bm{p}_0^{}$},
\begin{subequations}\label{eq:cpmotion}
\begin{align}
\ddt{\bm{x}}&=\,\bm{v}(\bm{x},\bm{p})\,,&\bm{x}(0)=\bm{x}_0^{}\,,\\
\ddt{\bm{p}}&=\,\bm{0}\,,&\bm{p}(0)=\bm{p}_0^{}\,.
\end{align}
\end{subequations}
These equations can be rewritten in the space~+~parameter coordinates by introducing $\bm{m}=[\bm{m^x_{}},\bm{m^p_{}}]^T_{}=[\bm{x},\bm{p}]^T_{}$ and the drift \mbox{$\bm{f}=[\bm{v},\bm{0}]^T_{}$}
\begin{align}\label{eq:aggrmotion}
\ddt{\bm{m}}&=\,\bm{f}(\bm{m})\,,&\bm{m}(0)=\bm{m}_0^{}\,,
\end{align}
Let us note that Eq.~\eqref{eq:cpmotion}, or its compact version Eq.~\eqref{eq:aggrmotion}, corresponds to the very specific case of constant parameters. While this in general true, there can be situations where this assumption is not valid. For example, as the particle moves in an inhomogeneous fluid, the physical parameters can have an implicit dependence on the particle position and vary during the motion of the particle along the trajectory, it results thus \mbox{$\bm{p}=\bm{p}(\bm{x})$}.

The essence of the method used by Yang and Vitali is to consider the dynamics of Eq.~\eqref{eq:aggrmotion2} for two perturbations with initial distance \mbox{$\bm{u}(0)=\bm{u}_0^{}$}, namely
\begin{align}\label{eq:aggrmotion2}
\ddt{\bm{m}_-^{}}&=\,\bm{f}(\bm{m}_-^{})\,,&\bm{m}_{-}^{}(0)=\bm{m}_0^{}-\tfrac{1}{2}\bm{u}_0^{}\,,\\
\ddt{\bm{m}_+^{}}&=\,\bm{f}(\bm{m}_+^{})\,,&\bm{m}_{+}^{}(0)=\bm{m}_0^{}+\tfrac{1}{2}\bm{u}_0^{}\,.
\end{align}
where \mbox{$\bm{u}_0^{}=\mathrm{BW}(0)\bm{\hat{y}}+\bm{u^p}_0$} corresponds to the initial value of the bandwidth $\mathrm{BW}$ along the $y$-direction plus the parameters' perturbation $\bm{u^p}_0$ with \mbox{$\bm{u^p_0}\cdot\bm{\hat y}=0$}.
Half-summing and subtracting equations \eqref{eq:aggrmotion2}, expanding up to the first order $\bm{f}$ in the small perturbation $\bm{u}$ around $\bm{m}$, and resulting in this case \mbox{$\bm{m}\simeq\tfrac{1}{2}(\bm{m}_+^{}+\bm{m}_-^{})$} and \mbox{$\bm{u}\simeq\bm{m}_+^{}-\bm{m}_-^{}$}, Eqs.~\eqref{eq:aggrmotion2} result equivalent to
\begin{subequations}\label{eq:lyapunov}
\begin{align}
\label{eq:lyapunov1}\ddt{\bm{m}}&=\,\bm{f}(\bm{m})\,,&\bm{m}(0)=\bm{m}_0^{}\,,\\
\label{eq:lyapunov2}\ddt{\bm{u}}&=\,\bm{J}(\bm{m})\cdot\bm{u}\,,&\bm{u}(0)=\bm{u}_0^{}\,.
\end{align}
\end{subequations}
where \mbox{$\bm{J}(\bm{m})=\bm{\nabla}\bm{f}(\bm{m})$} is the Jacobian of the drift computed in the correspondence of $\bm{m}$.
Noting that Eq.~\eqref{eq:lyapunov2} can be rewritten in Einstein notation (\mbox{$h=1...N+3$})
\begin{align}
\ddt{u^h_{}}&=\,\partial_l^{}f^h_{}(\bm{m})\,u^l_{}\,,&u^h(0)=u_0^h\,,
\end{align}
introducing the Kronecker product of the perturbation components \mbox{$s^{hk}_{}=u^h_{}u^k_{}$}, and considering that its time-derivative is
\begin{equation}
\ddt{s^{hk}_{}}=\ddt{u^h_{}}\,u^k_{}+u^h_{}\,\ddt{u^k_{}}\,,
\end{equation}
in place of the Eqs.~\eqref{eq:lyapunov}, the system
\begin{subequations}\label{eq:mcd}
\begin{align}
\label{eq:mcd1}\ddt{\bm{m}}&=\,\bm{f}(\bm{m})\,,&\bm{m}(0)=\bm{m}_0^{}\,,\\
\label{eq:mcd2}\ddt{\bm{s}}&=\,\bm{J}(\bm{m})\cdot\bm{s}+\bm{s}\cdot\bm{J}^T_{}(\bm{m})\,,&\bm{s}(0)=\bm{s}_0^{}\,,
\end{align}
\end{subequations}
can be considered, and in which \mbox{$\bm{s}_0^{}=\bm{u}_0^{}\bm{u}_0^T$} is the initial value for $\bm{s}$.
Note that, the covariance matrix is \mbox{$(3+N)\times(3+N)$} symmetric matrix 
\begin{equation}
\bm{s}=\left(\begin{array}{cc}
\bm{s^{xx}_{}} & (\bm{s^{xp}_{}})^T \\
\bm{s^{xp}_{}} & \bm{s^{pp}_{}}
\end{array}
\right)
\end{equation}
where $\bm{s^{xx}_{}}$ is a $3\times 3$ matrix, $\bm{s^{pp}_{}}$ is a $N\times N$ matrix, and $\bm{s^{xp}_{}}$ is a $N\times 3$ matrix.
Note that for the parameter components results \mbox{$\tfrac{\mathrm{d}}{\mathrm{d}t}\bm{m^p_{}}=\bm{0}$} and \mbox{$\tfrac{\mathrm{d}}{\mathrm{d}t}\bm{s^{pp}_{}}=\bm{0}$}.

Equations~\eqref{eq:mcd} are the multivariate and diffusion-less form of equations used for studying dispersion in~\cite{garofalo2014}, and extended to microparticle populations with arbitrary statistics, that is equivalent to large perturbations, in~\cite{garofalo2017,garofalo2018}. Equations~\eqref{eq:mcd} have been addressed as mean-and-covariance dynamics in~\cite{garofalo2018}, where also their connection with stochastic linearization methods has been recognized.

\subsection{Indicators}
The knowledge of the mean and the covariance by solution of Eqs.~\eqref{eq:mcd} enables for the approximation of the dynamics for the probability density function (PDF)
\begin{equation}\label{eq:pdf}
\rho(\bm{q},t)=\mathcal{N}[\,\bm{q}\,|\,\bm{m}(t),\bm{s}(t)\,]\,,
\end{equation}
where $\mathcal{N}(\bm{q}\,|\,\bm{m},\bm{s})$ is a multivariate normal distribution with mean $\bm{m}$ and covariance matrix $\bm{s}$.
Most relevant for the present discussion is the spatial marginal
\begin{equation}\label{eq:pdf}
\rho^{\mathrm{spatial}}(\bm{x},t)=\mathcal{N}[\,\bm{x}\,|\,\bm{m^x_{}}(t),\bm{s^{xx}_{}}(t)\,]\,,
\end{equation}
that with a suitable parametrization, e.g. in the $x$-direction, enables for the introduction of indicators similar to those addressed by Yang and used by Vitali.

\paragraph*{Bandwidth.}~From the derivation of the dynamics of the covariance $\bm{s}$ by the dynamics of $\bm{u}$ in the tangent space, we infer that the bandwidth $\mathrm{BW}$ is comparable to the square-root of the variance in the $y$-direction
\begin{equation}
\mathrm{BW}(\xi) = |[\bm{m}_+^{}(\xi)-\bm{m}_-^{}(\xi)]\cdot\bm{\hat y}| \simeq \sqrt{s^{yy}_{}}(\xi)\,,
\end{equation}
where $\xi$ is a generic coordinate corresponding to the ``section'' where the comparison is performed.
\paragraph*{Separation Efficiencies.}~A quantity similar to the Yang's ideal separation efficiency $\mathrm{ISE}$ can be derived from the statistical method by computing the integral along the $y$-direction of the product of two spatial marginals at a given section $\xi$~\cite{garofalo2014_2}.
The inverse of the separation resolution is
\begin{align}
\mathrm{SR}^{-1}_{}(\xi)&=\,\int_0^H\rho_1^\mathrm{spatial}(\xi,y)\,\rho_2^\mathrm{spatial}(\xi,y)\,\mathrm{d}y\simeq\nonumber\\
\label{eq:sr}&\simeq\,\int_{-\infty}^{\infty}\rho_1^\mathrm{spatial}(\xi,y)\,\rho_2^\mathrm{spatial}(\xi,y)\,\mathrm{d}y\,,
\end{align}
and this integral admits an analytical solution in the case when the marginals are gaussians
\begin{equation}
\mathrm{SR}^{-1}_{}(\xi)=\frac{\exp\left\{-\frac{[m_1^y(\xi)-m_2^y(\xi)]^2}{2[s^{yy}_1(\xi)+s^{yy}_2(\xi)]}\right\}}{\sqrt{2\,\pi\,[s^{yy}_1(\xi)+s^{yy}_2(\xi)]}}\,.
\end{equation}
Note the similitude between the square-root of the exponential argument, here named resolution index $\mathrm{RI}$, and the Yang's $\mathrm{ISE}$
\begin{align}
\mathrm{RI}(\xi)&=\,\frac{|m_1^y(\xi)-m_2^y(\xi)|}{\sqrt{2}[s^{yy}_1(\xi)+s^{yy}_2(\xi)]^{1/2}_{}}\,,\\
\mathrm{ISE}(\xi)&=\,2\,\frac{|\braket{y_1^{}}(\xi)-\braket{y_2^{}}(\xi)|}{\mathrm{BW}_1^{}(\xi)+\mathrm{BW}_2^{}(\xi)}\,,
\end{align}
where \mbox{$\braket{y}=\tfrac{1}{2}(\bm{m}_+^{}+\bm{m}_-^{})\cdot\bm{\hat y}$}.
Finally, the Yang's separation efficiency $\mathrm{SE}$ is essemptially the ideal separation efficiency $\mathrm{ISE}$ with a radius correction. There is no equivalent/similar quantity in the statistical approach proposed by Garofalo~\cite{garofalo2018}.
Furthermore, some issues arise when this quantity is analyzed as
the separation efficiency $\mathrm{SE}$ is defined only for particles with the same radii, and this induces some limitations in the method used by Yang and Vitali.
Conversely, the statistics-based method considers the dependence of the dispersion properties on the coordinates and parameters on equal footing, embedding all these contributions in the dynamics of $\bm{s}$ and, as a consequence, their impact is encoded in both $\mathrm{SR}$ and $\mathrm{RI}$.

\section{Results}
In order to show the quantitative comparison of the two approaches while performing effortless analytical calculations, let us consider the following prototypical model for acoustophoresis reported in Section~III.A of~\cite{garofalo2018}
\begin{equation}\label{eq:toymodel}
v(y,r)=\alpha\,r^2\,\sin(2\,\pi\,y)\,,
\end{equation}
for which the mean-and-covariance dynamics results
\begin{subequations}\label{eq:meancovmodel}
\begin{align}
\ddt{{m}^y_{}}&=\,\alpha\,(m_0^r)^2\,\sin(2\pi\,m^y_{})\,,\\
\ddt{{s}^{yr}_{}}&=\,\alpha\,\left[2\pi\,(m_0^r)^2\,\cos(2\pi\,m^y_{})\,s^{yr}_{}+\nonumber\right.\\
&\left.+\,2\,m_0^r\,\sin(2\pi\,m^y_{})\,s^{rr}_0\right]\,,\\
\ddt{{s}^{yy}_{}}&=\,\alpha\,\left[4\pi\,(m_0^r)^2\,\cos(2\pi\,m^y_{})\,s^{yy}_{}+\nonumber\right.\\
&\left.+4\,m_0^r\,\sin(2\pi\,m^y_{})\,s^{yr}_{}\right]\,,
\end{align}
\end{subequations}
where $m_0^r$ and $s_0^{rr}$ are the mean radius and its variance, respectively. The initial conditions for this system are chosen as to compare the bandwidth areas of and the dispersion bands in the Garofalo's method (we consider for all the cases \mbox{$s^{yr}_{}(0)=0$}, and \mbox{$\alpha=1$} except when the comparison between the indicators is considered).

\begin{figure}[!!t]
%\fbox{\begin{picture}(226,148)
%\fbox{
\begin{picture}(226,405)
\put(-40,-15){\includegraphics[width=10cm]{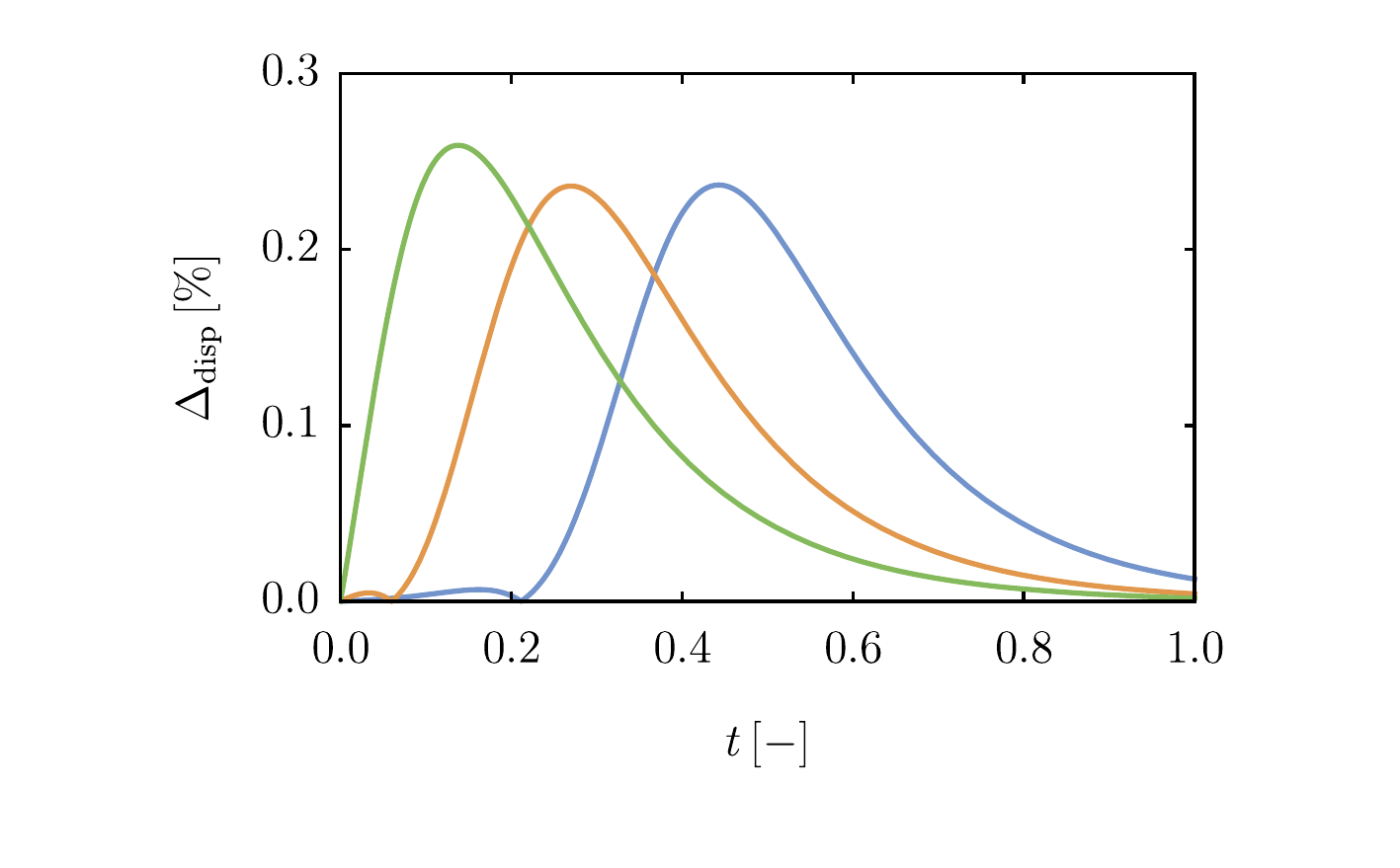}}
\put(-40,115){\includegraphics[width=10cm]{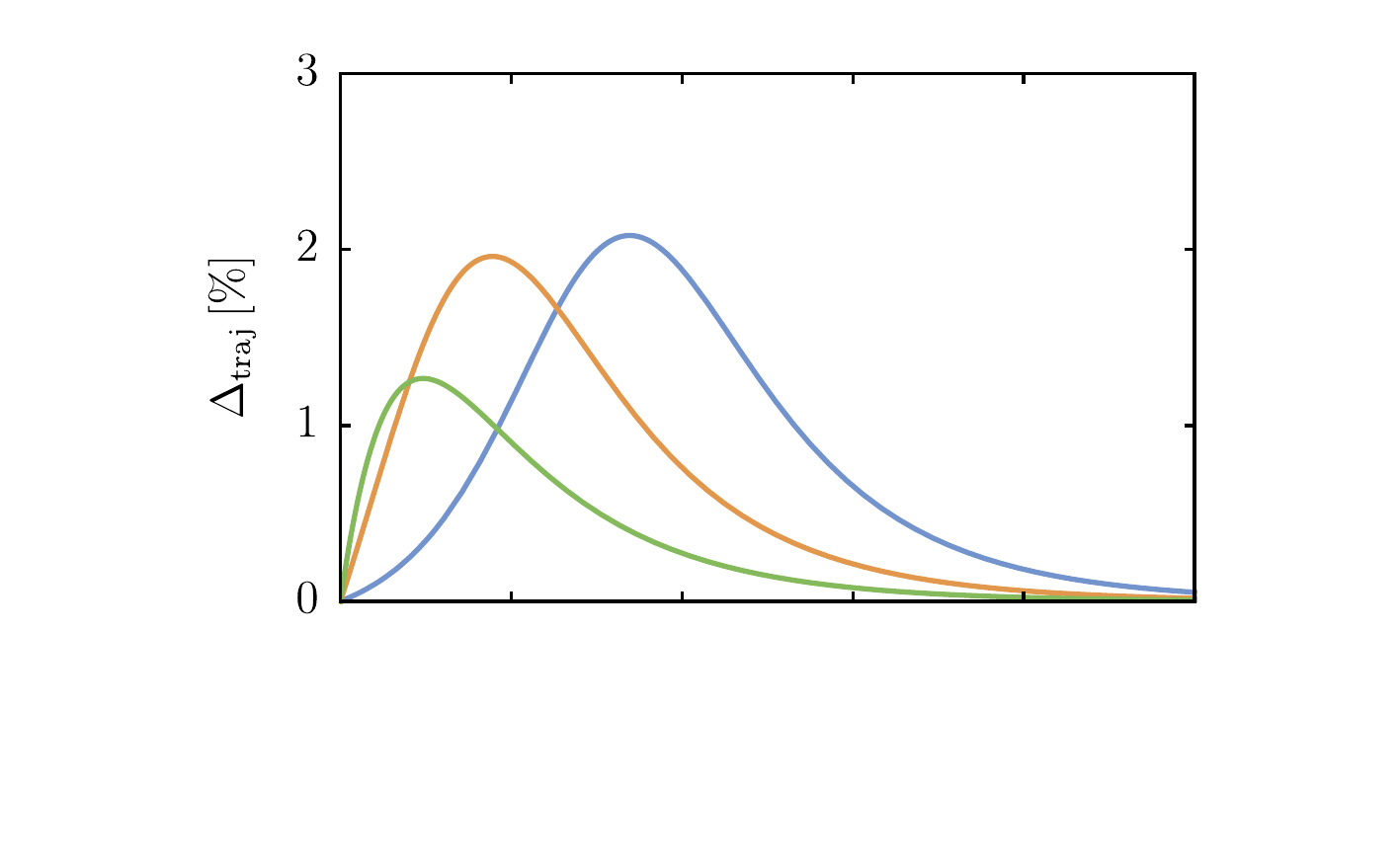}}
\put(-40,245){\includegraphics[width=10cm]{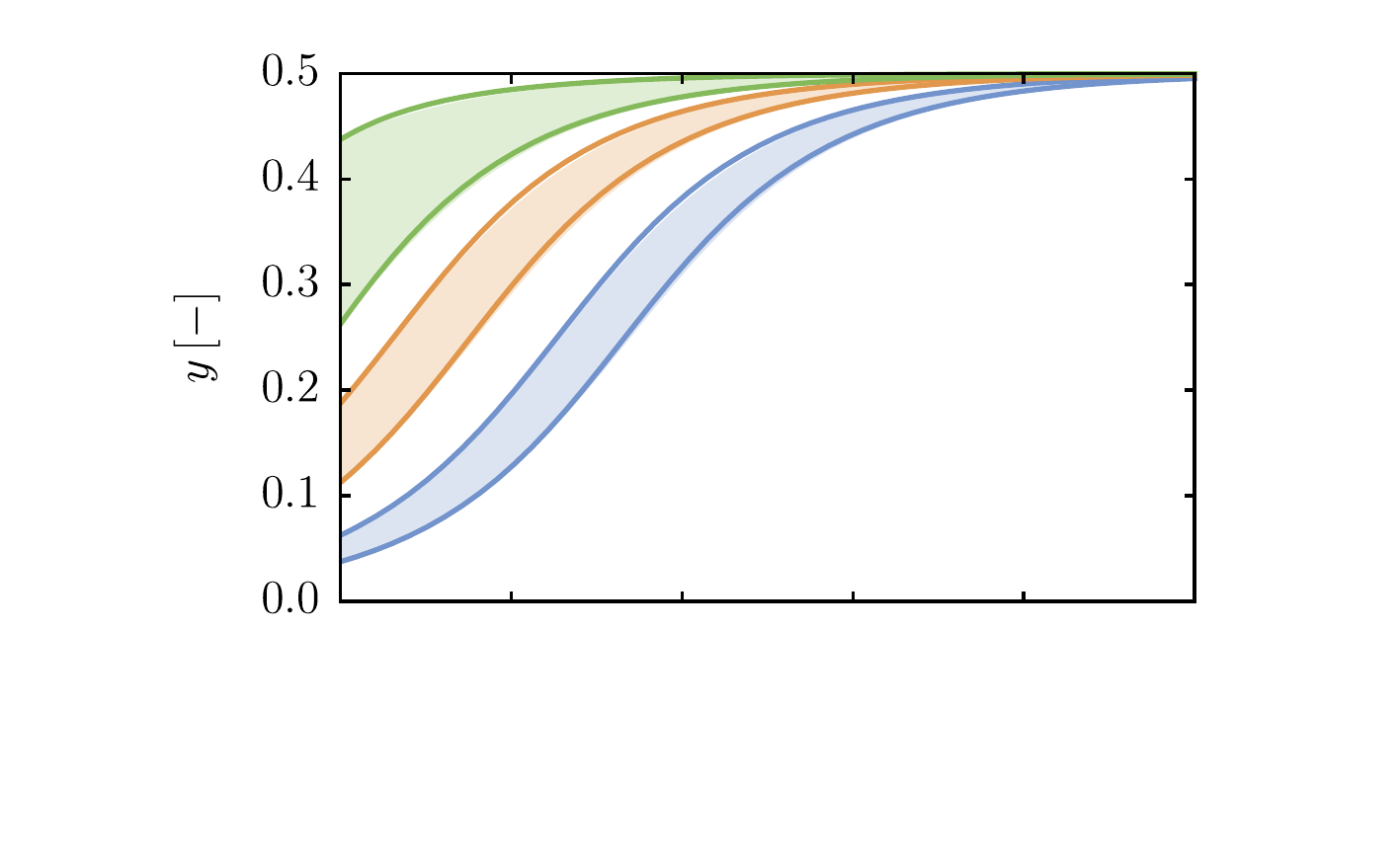}}
\put(187,396){(A)}
\put(187,266){(B)}
\put(187,136){(C)}
\end{picture}
%}
\caption{(Color Online) Results for the model Eq.~\eqref{eq:toymodel}. (A)~Bandwidth areas (colored areas) and dispersion bands \mbox{$m^y_{}\pm\tfrac{1}{2}\sqrt{s^{yy}_{}}$} (lines) as function of the time $t$.
(B)~Distance between the average trajectories $\mathrm{\Delta}_\mathrm{traj}^{}$ and (C)~distance between the dispersion characteristics $\mathrm{\Delta}_\mathrm{disp}^{}$ as function of the time $t$ corresponding to the cases in panel~A.
The initial conditions are:  \mbox{$y_0^{}=0.05$} (blue), \mbox{$y_0^{}=0.15$} (orange), and \mbox{$y_0^{}=0.35$} (green). For all of the simulations it has \mbox{$m_0^r=1$}, \mbox{$s_0^{rr}=0$}, \mbox{$s_0^{yr}=0$}, and \mbox{$\sqrt{s^{yy}_0}=\mathrm{BW}(0)=50\%\,y_0^{}$}.
}
\label{fig:comparison}
\end{figure}
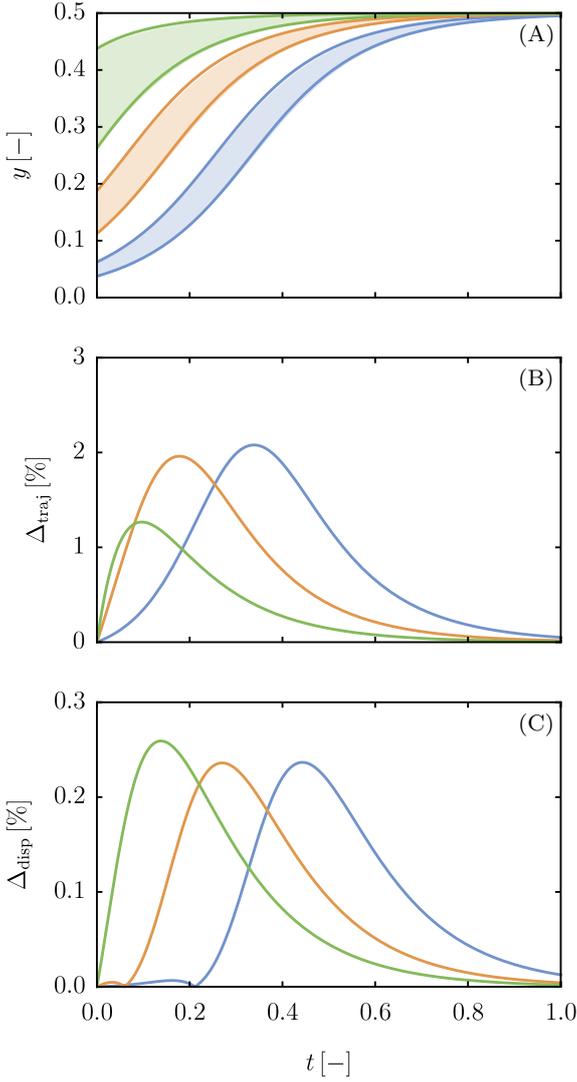

Figure~\ref{fig:comparison}~(A) shows the comparison between the two methods for different values of the inlet condition $y_0^{}$, a relative initial perturbation equals to $\mathrm{BW}(0)=\sqrt{s^{yy}_0}=50\%\,y_0^{}$, radius \mbox{$m^r_{}=1$}, and no initial perturbation on the radius \mbox{$s^{rr}_{}=0$}.
Panel (A) reports the comparison between the bandwidth areas \mbox{$y\in[y_{-}^{},y_+^{}]$} and the dispersion bands \mbox{$m^y_{}\pm\tfrac{1}{2}\sqrt{s^{yy}_{}}$}.
It can see the results for the two methods almost coincide, and for quantifying the small differences we introduce the distances
\begin{subequations}
\begin{align}
\label{eq:deltatraj}\mathrm{\Delta}_\mathrm{traj}^{}(t)&=\,\frac{|m^y_{}(t)-\braket{y}(t)|}{m^y_{}(t)}\,,\\
\label{eq:deltadisp}\mathrm{\Delta}_\mathrm{disp}^{}(t)&=\,\frac{|\sqrt{s^{yy}_{}(t)}-\mathrm{BW}(t)|}{1+\sqrt{s^{yy}_{}(t)}}\,,
\end{align}
\end{subequations}
where the factor $1$ was included in the denominator to prevent divergent values for $\mathrm{\Delta}_\mathrm{disp}^{}$ since the variance approaches zero for large $t$.
These quantities are reported in Fig.~\ref{fig:comparison}(B) and (C), respectively.
As it can see from a practical point-of-view and in small perturbation cases, the two methods are equivalent with a maximum distance for the trajectories less than $3\%$ and less than $0.3\%$ for the maximum distance in the dispersion.

\begin{figure}[!!t]
%\fbox{\begin{picture}(226,148)
%\fbox{
\begin{picture}(226,405)
\put(-40,-15){\includegraphics[width=10cm]{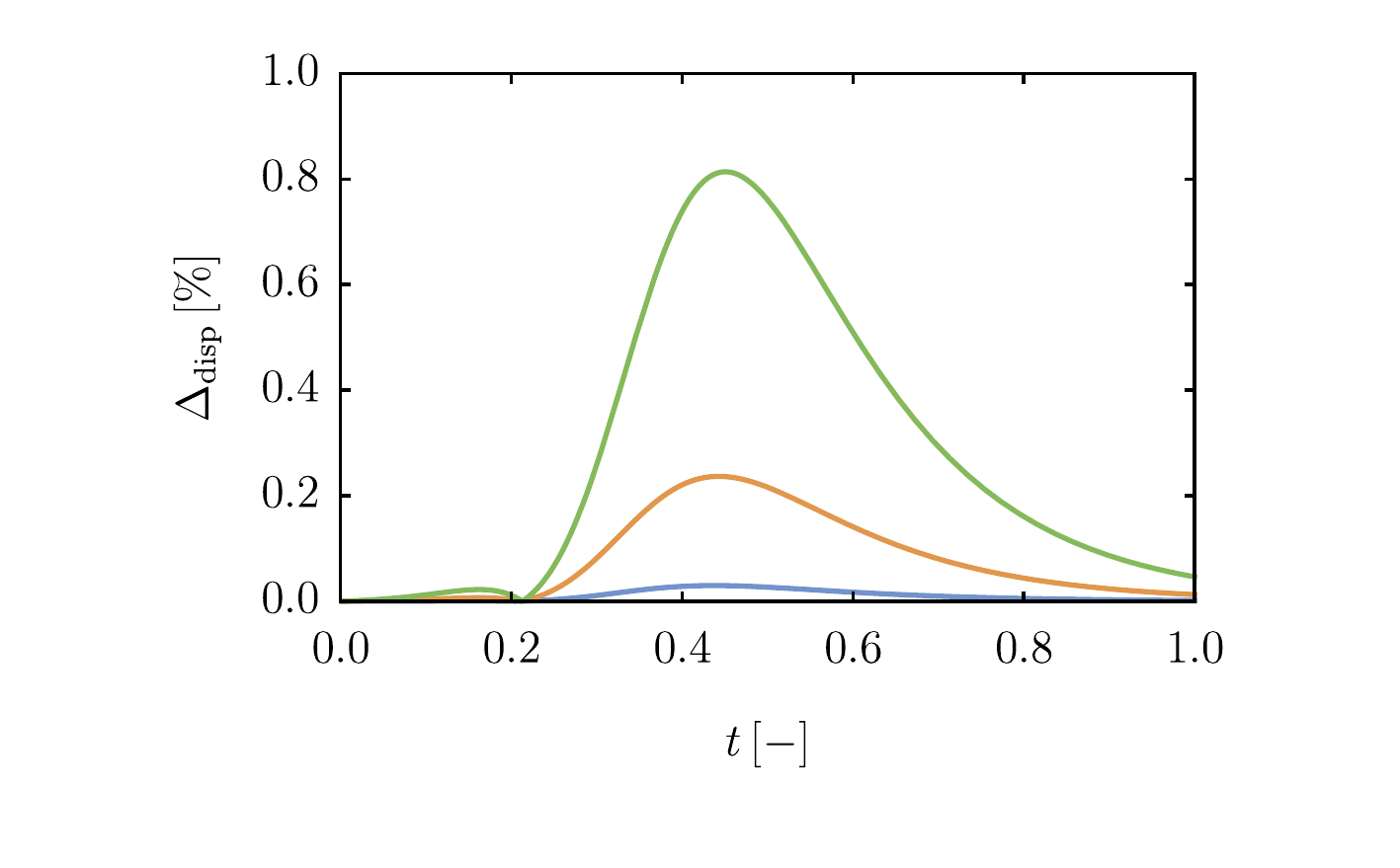}}
\put(-40,115){\includegraphics[width=10cm]{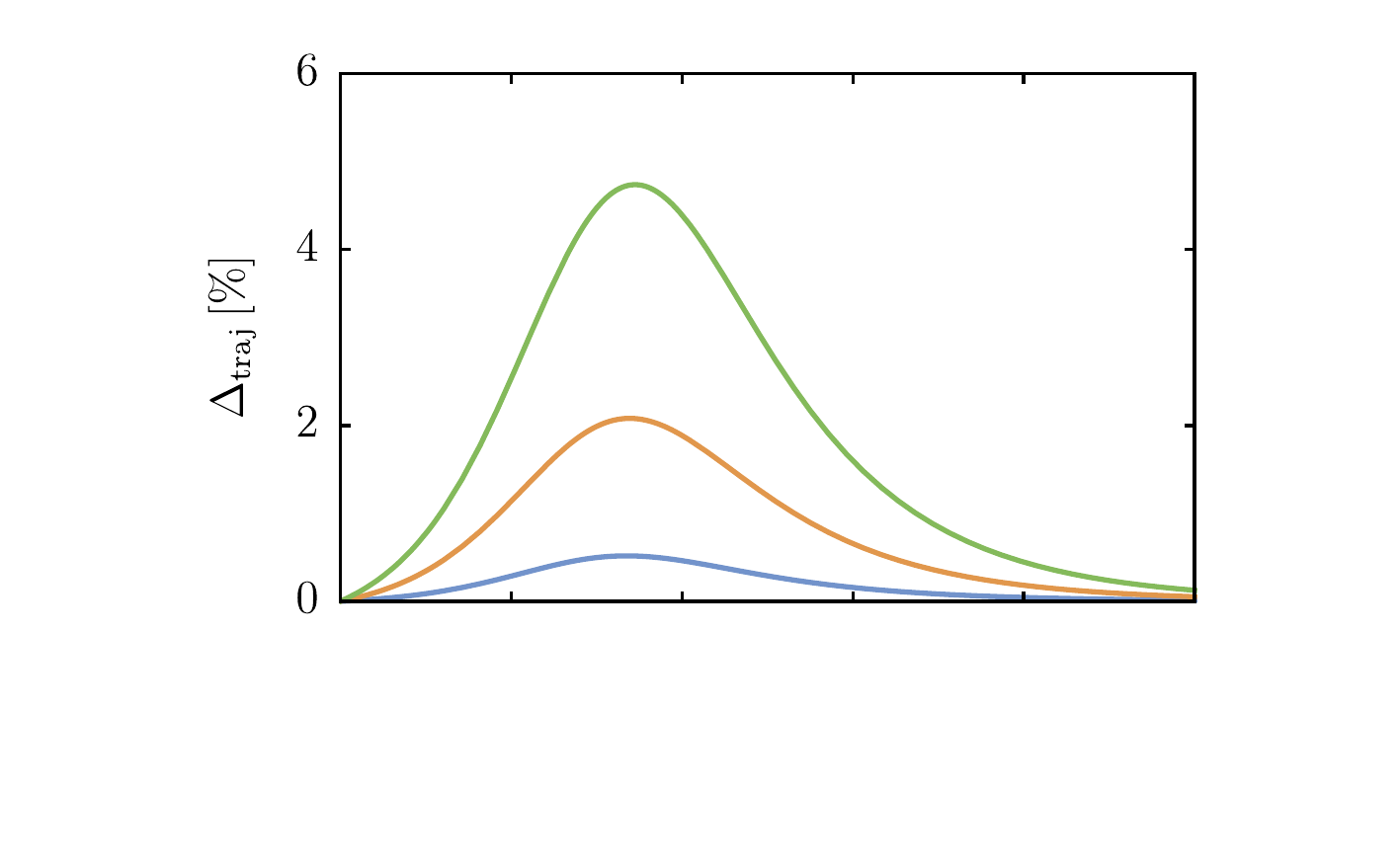}}
\put(-40,245){\includegraphics[width=10cm]{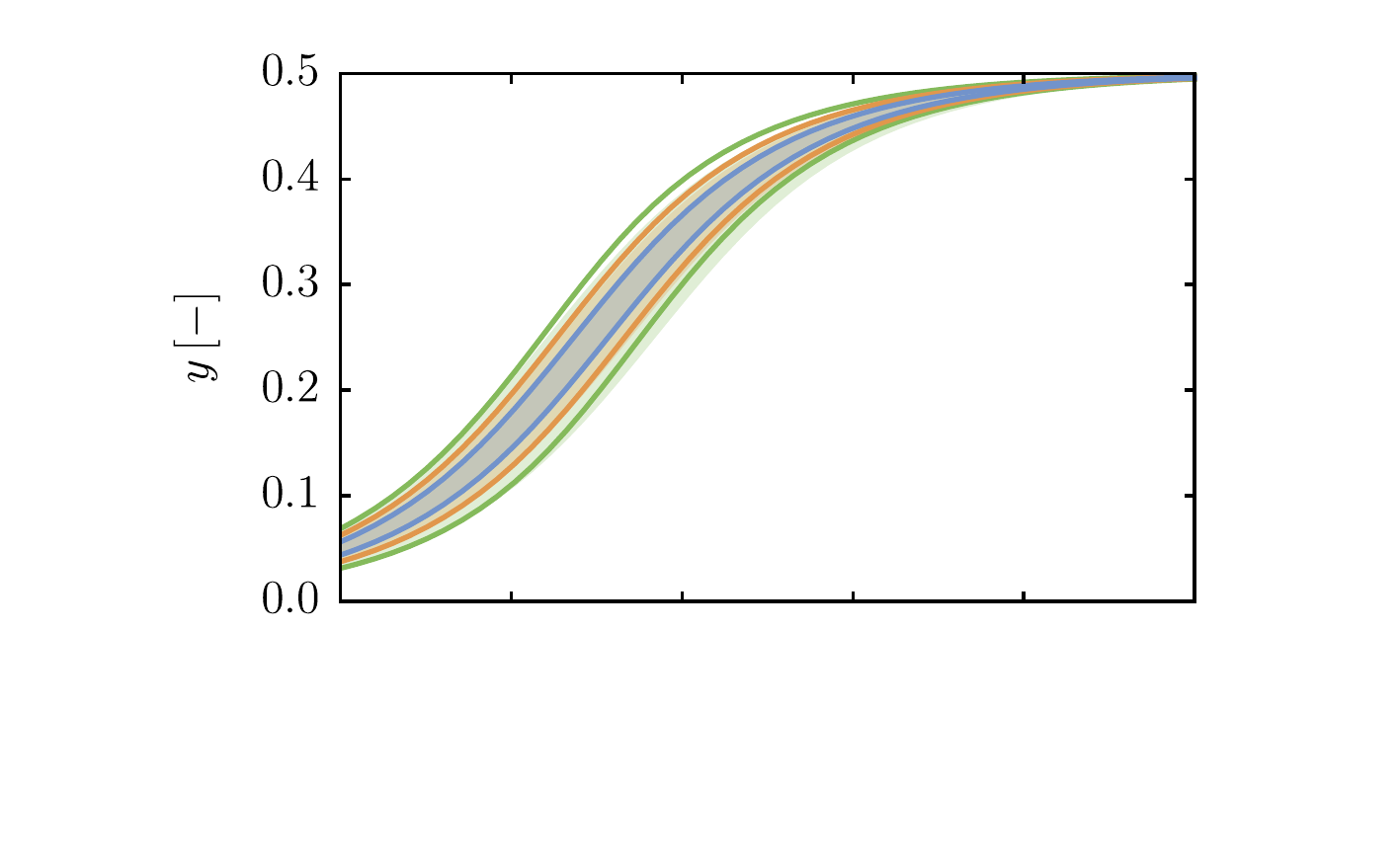}}
\put(187,396){(A)}
\put(187,266){(B)}
\put(187,136){(C)}
\end{picture}
%}
\caption{(Color Online) Results for the model Eq.~\eqref{eq:toymodel}. (A)~Bandwidth areas (colored areas) and dispersion bands \mbox{$m^y_{}\pm\tfrac{1}{2}\sqrt{s^{yy}_{}}$} (lines) as function of the time $t$.
(B)~Distance between the average trajectories $\mathrm{\Delta}_\mathrm{traj}^{}$ and (C)~distance between the dispersion characteristics $\mathrm{\Delta}_\mathrm{disp}^{}$ as function of the time $t$ corresponding to the cases in panel~A.
The initial conditions are for \mbox{$y_0^{}=0.05$} and:
\mbox{$\sqrt{s^{yy}_0}=25\%\,y_0^{}$} (blue),
\mbox{$\sqrt{s^{yy}_0}=50\%\,y_0^{}$} (orange),
\mbox{$\sqrt{s^{yy}_0}=75\%\,y_0^{}$} (green).
For all of the simulations it has \mbox{$m_0^r=1$} and \mbox{$s_0^{rr}=0$}.
}
\label{fig:comparison2}
\end{figure}

Figure~\ref{fig:comparison2} shows an analogous comparison. In this cases the different curves are for the same inlet condition $y_0^{}$ and different values of the initial dispersion or bandwidth.
It is noticeable that for bandwidth less than $100\%\,y_0^{}$ the two methods give yet very close results, i.e. within $6\%$ in the trajectory distance $\mathrm{\Delta}_\mathrm{traj}^{}$ and within $1\%$ in the dispersion distance $\mathrm{\Delta}_\mathrm{disp}^{}$.

Let us consider now the case when the initial perturbation occurs in the spatial and radius components, that is when $\sqrt{s^{rr}_0}\neq 0$.
In this case one can force the methodology proposed by Yang et al. by considering the perturbation \mbox{$\bm{u}_0^{}=\sqrt{s^{xx}_0}\bm{\hat y}+\sqrt{s^{rr}_0}\bm{\hat r}$}, where both the $y$-direction and $r$-direction are meant in the coordinate+parameter space.
This specific perturbation was chosen by considering the ``best'' and ``worst'' cases: the trajectory starting closer to the wall features a radius smaller than the average, and the trajectory starting far from the wall features a radius bigger than the average.
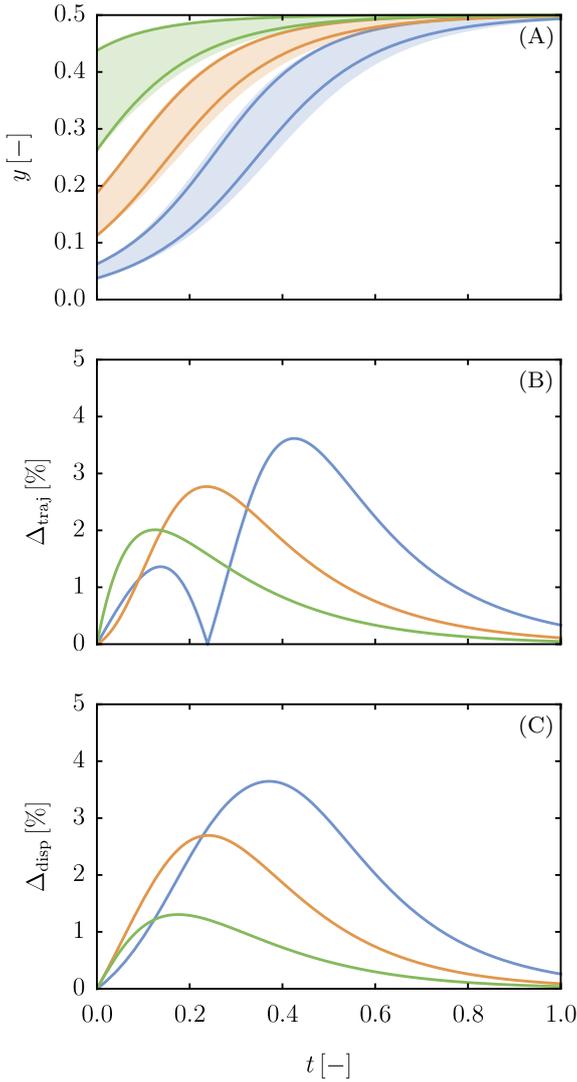
\begin{figure}[!!t]
%\fbox{\begin{picture}(226,148)
%\fbox{
\begin{picture}(226,405)
\put(-40,-15){\includegraphics[width=10cm]{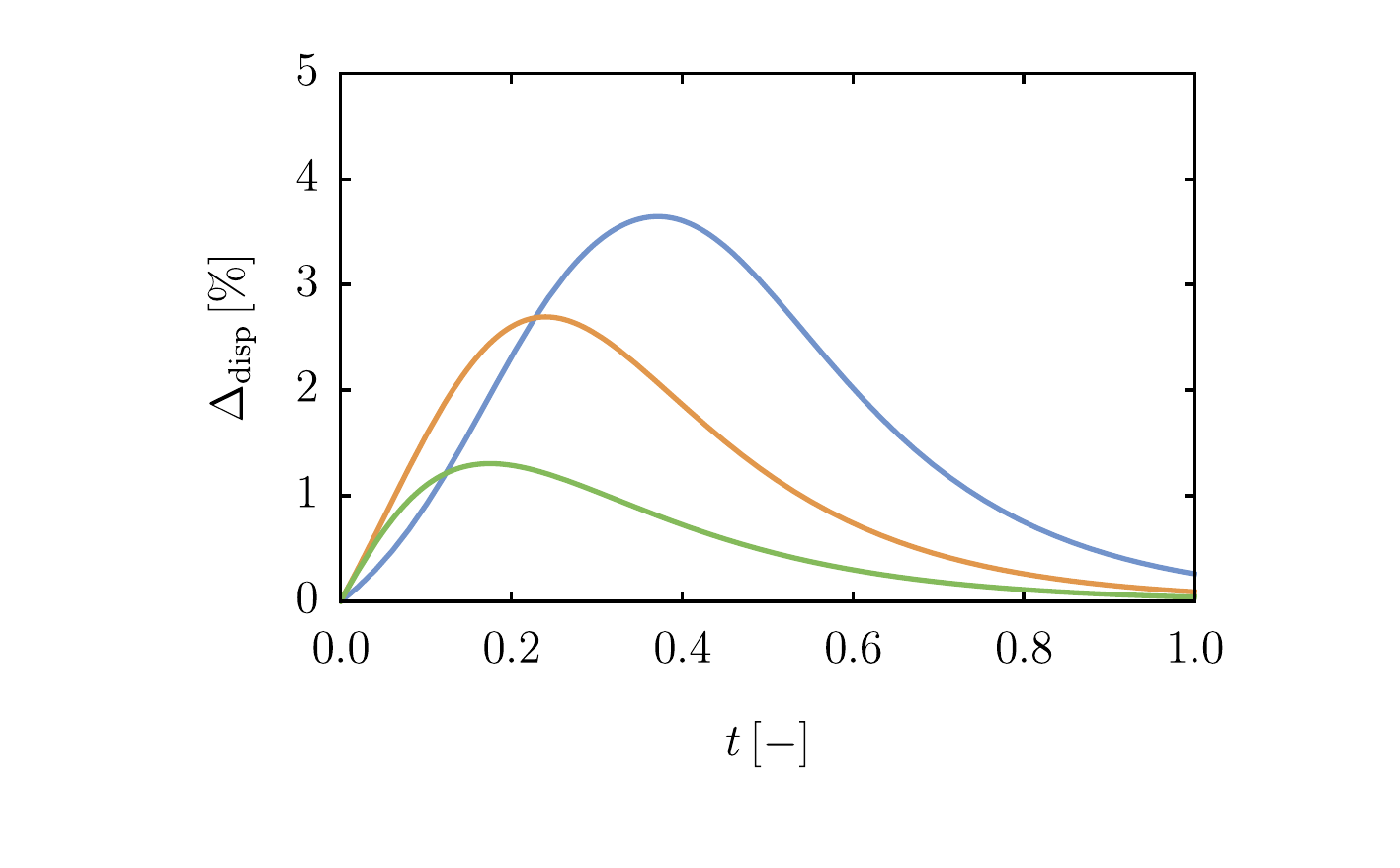}}
\put(-40,115){\includegraphics[width=10cm]{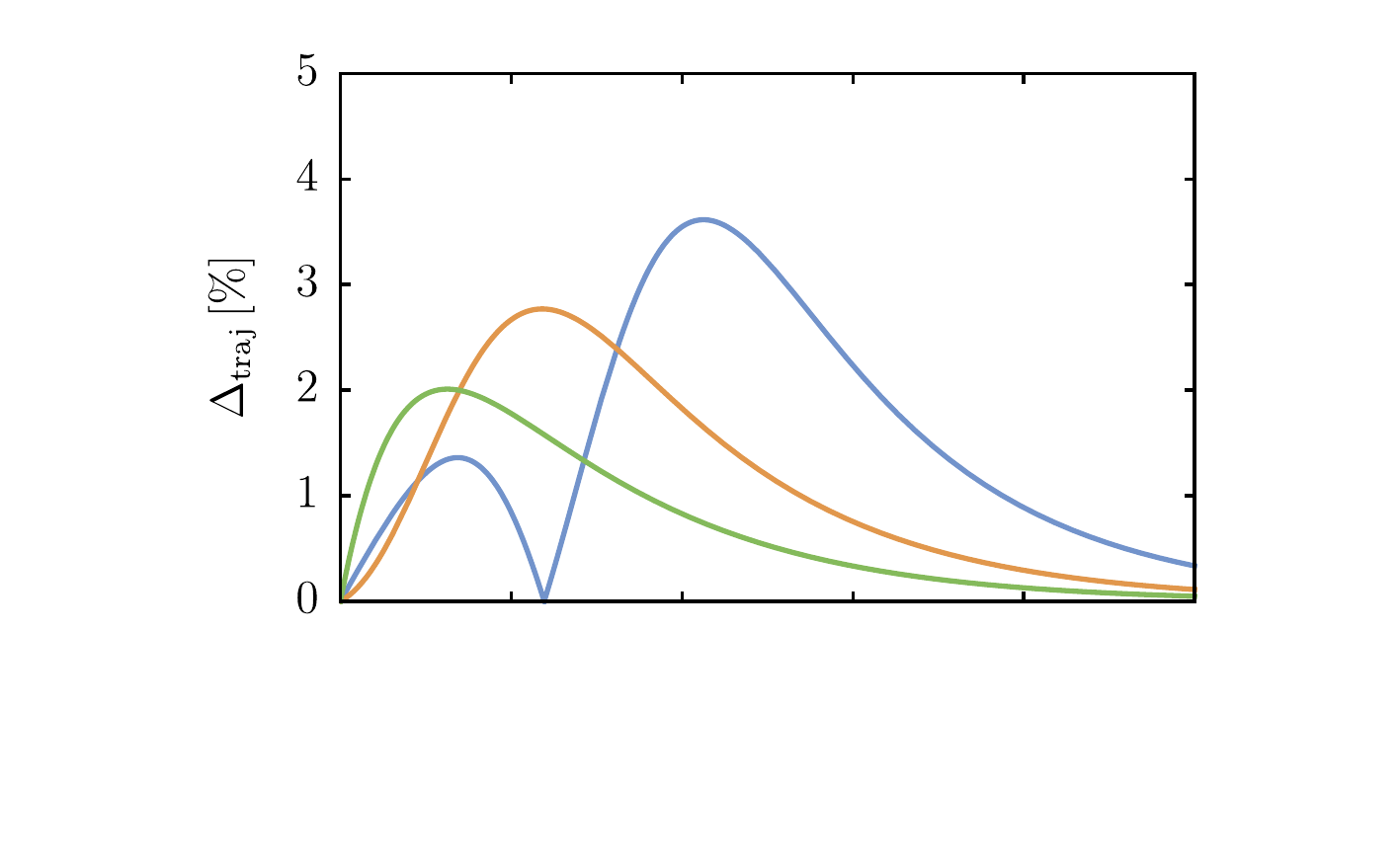}}
\put(-40,245){\includegraphics[width=10cm]{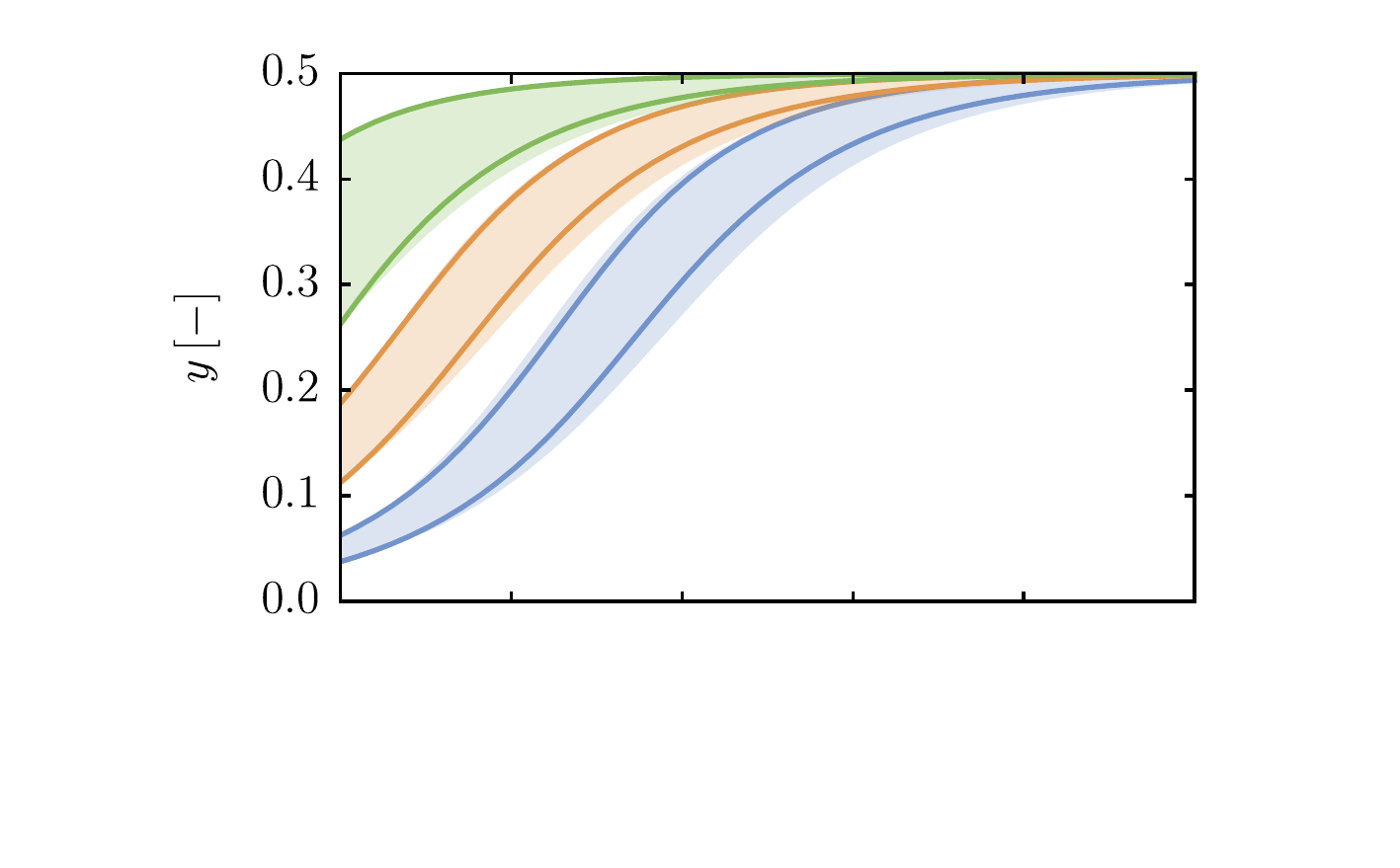}}
\put(187,396){(A)}
\put(187,266){(B)}
\put(187,136){(C)}
\end{picture}
%}
\caption{(Color Online) Results for the model Eq.~\eqref{eq:toymodel}. (A)~Bandwidth areas (colored areas) and dispersion bands \mbox{$m^y_{}\pm\tfrac{1}{2}\sqrt{s^{yy}_{}}$} (lines) as function of the time $t$.
(B)~Distance between the average trajectories $\mathrm{\Delta}_\mathrm{traj}^{}$ and (C)~distance between the dispersion characteristics $\mathrm{\Delta}_\mathrm{disp}^{}$ as function of the time $t$ corresponding to the cases in panel~A.
The initial conditions are:  \mbox{$y_0^{}=0.05$} (blue), \mbox{$y_0^{}=0.15$} (orange), and \mbox{$y_0^{}=0.35$} (green). For all of the simulations it has \mbox{$m_0^r=1$}, \mbox{$\sqrt{s_0^{rr}}=0.1$}, \mbox{$s_0^{yr}=0$}, and \mbox{$\sqrt{s^{yy}_0}=\mathrm{BW}(0)=50\%\,y_0^{}$}.
}
\label{fig:comparison3}
\end{figure}
Figure~\ref{fig:comparison3} shows the comparison between the two methods in terms of bandwidth areas and dispersion bands (panel~A) for different values of the inlet condition $y_0^{}$, an initial perturbation equals to $50\%y_0^{}$, radius \mbox{$m^r_{}=1$}, and initial perturbation for the radius equals \mbox{$\sqrt{s^{rr}_0}=0.1$}. In this plot is easily visible the fact that the two methods do not give exactly the same results, indeed observing the behaviors of $\mathrm{\Delta}_\mathrm{traj}^{}$ (panel~B) and $\mathrm{\Delta}_\mathrm{disp}^{}$ (panel~C) it can appreciate that the distances are larger than in the univariate cases, especially for that concerns the distance in the dispersion characteristics $\mathrm{\Delta}_\mathrm{disp}^{}$ (see Sec.~\ref{sec:discussion} for a discussion).

Finally, the comparison between the separation indicators is illustrated. Here the use of the parameter $\alpha$ is necessary as in the Yang's method the separation efficiency is defined for particle with the same radius, i.e. $R$ in Eq.~(13) in~\cite{yang2018}.
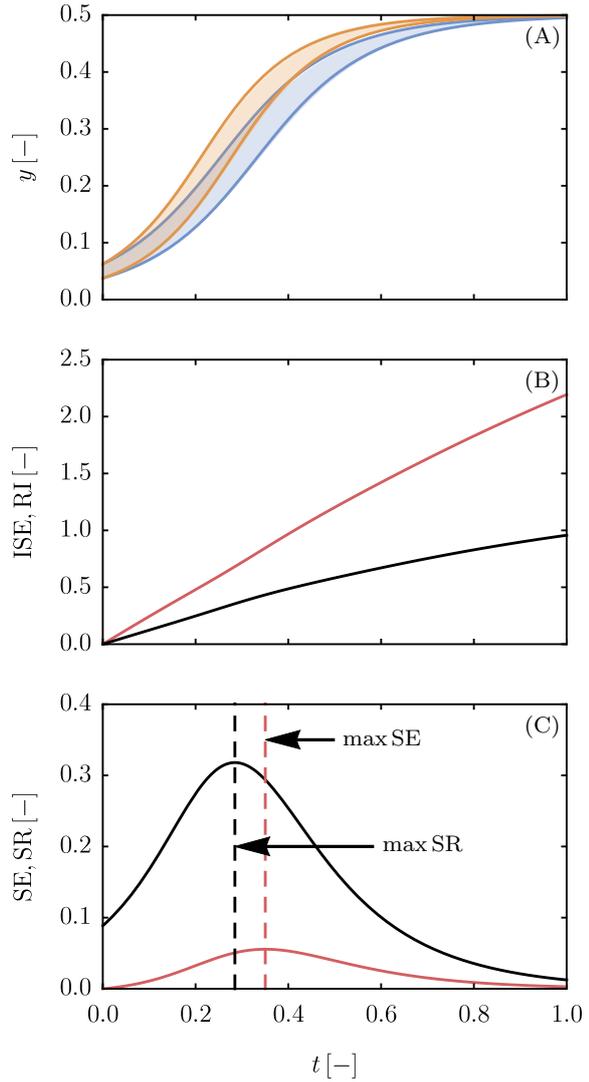
\begin{figure}[!!t]
%\fbox{\begin{picture}(226,148)
%\fbox{
%\begin{picture}(226,535)
\begin{picture}(226,405)
%\put(-40,-15){\includegraphics[width=10cm]{inverse_resolution.pdf}}
\put(-40,-15){\includegraphics[width=10cm]{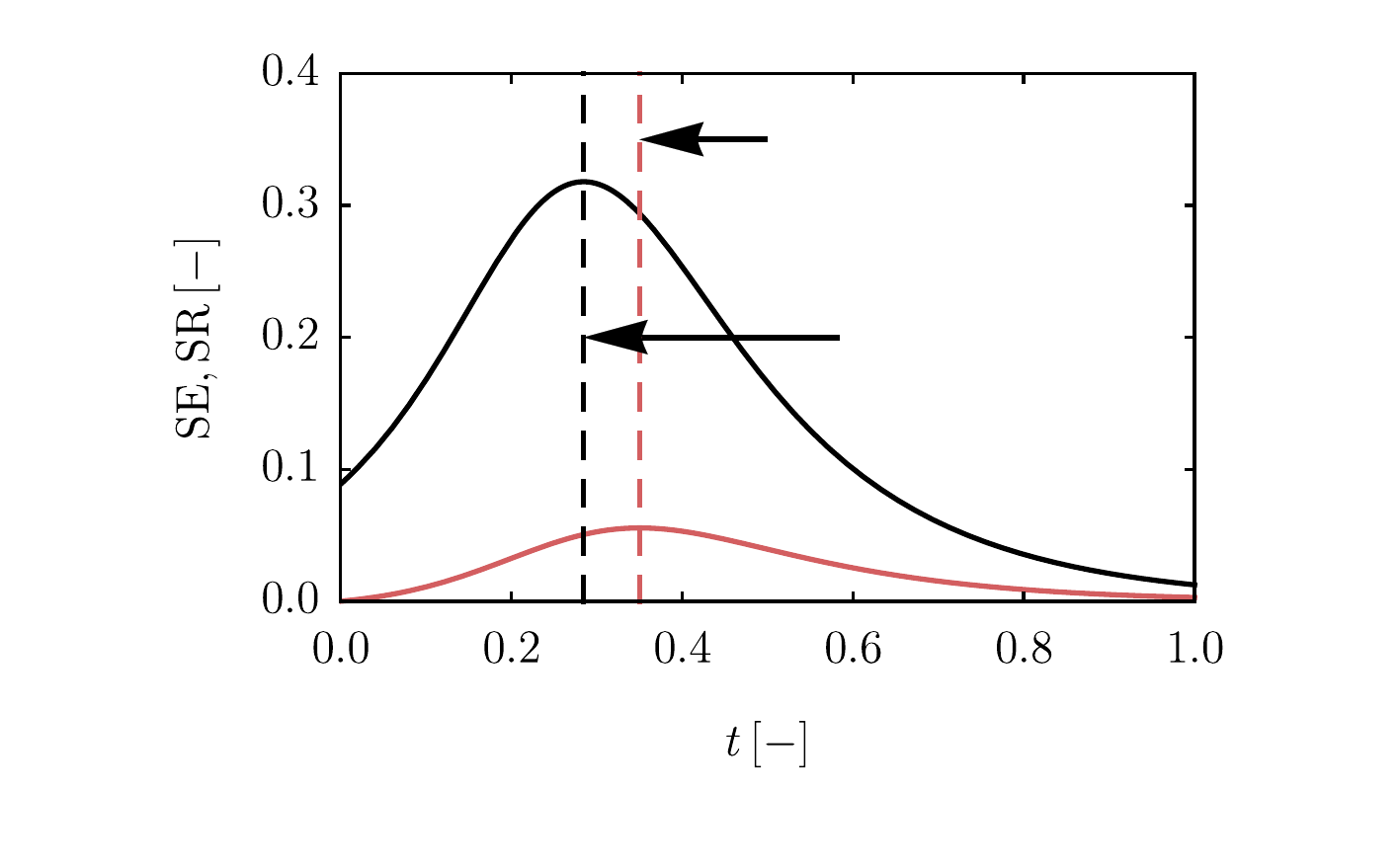}}
\put(-40,115){\includegraphics[width=10cm]{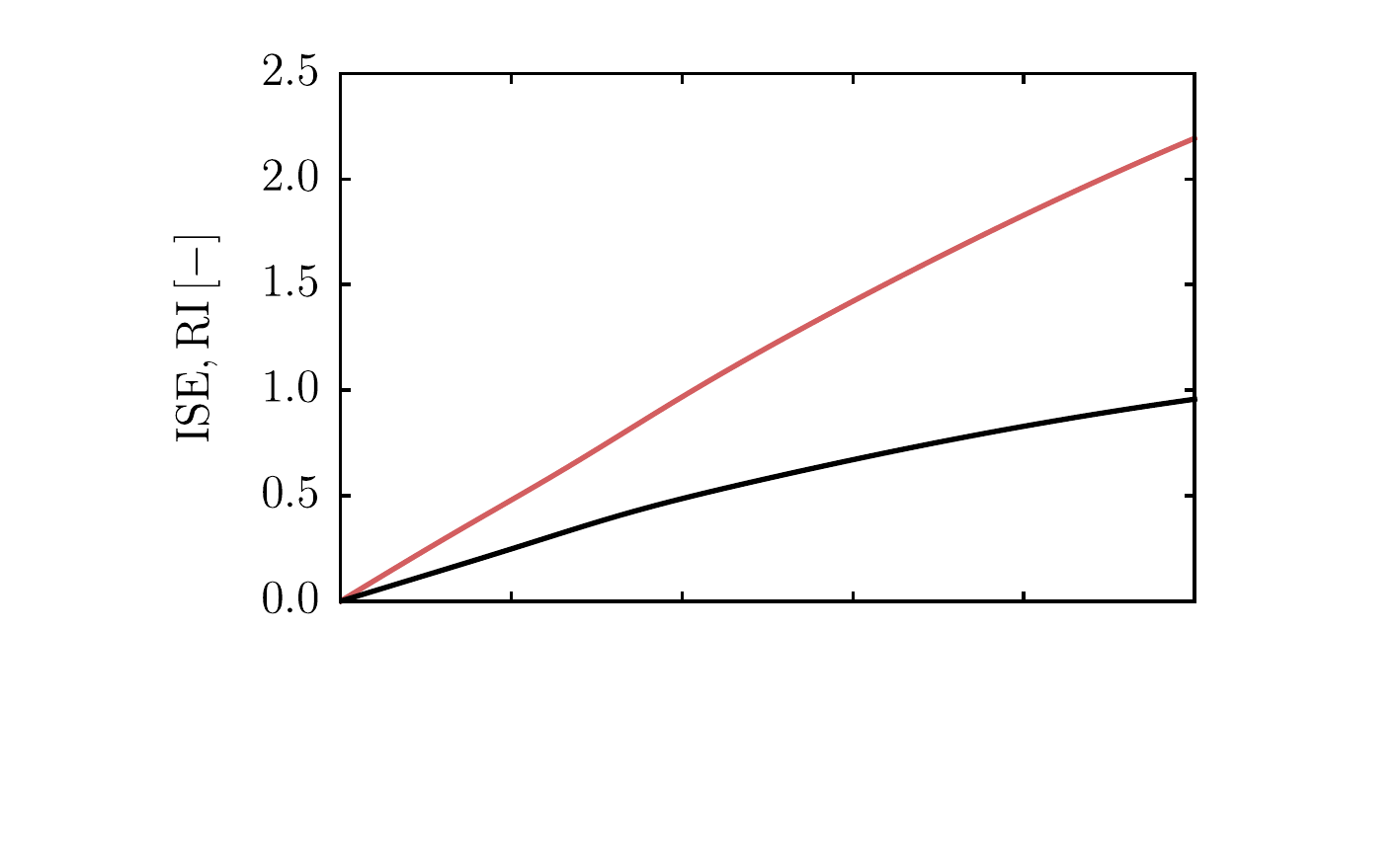}}
\put(-40,245){\includegraphics[width=10cm]{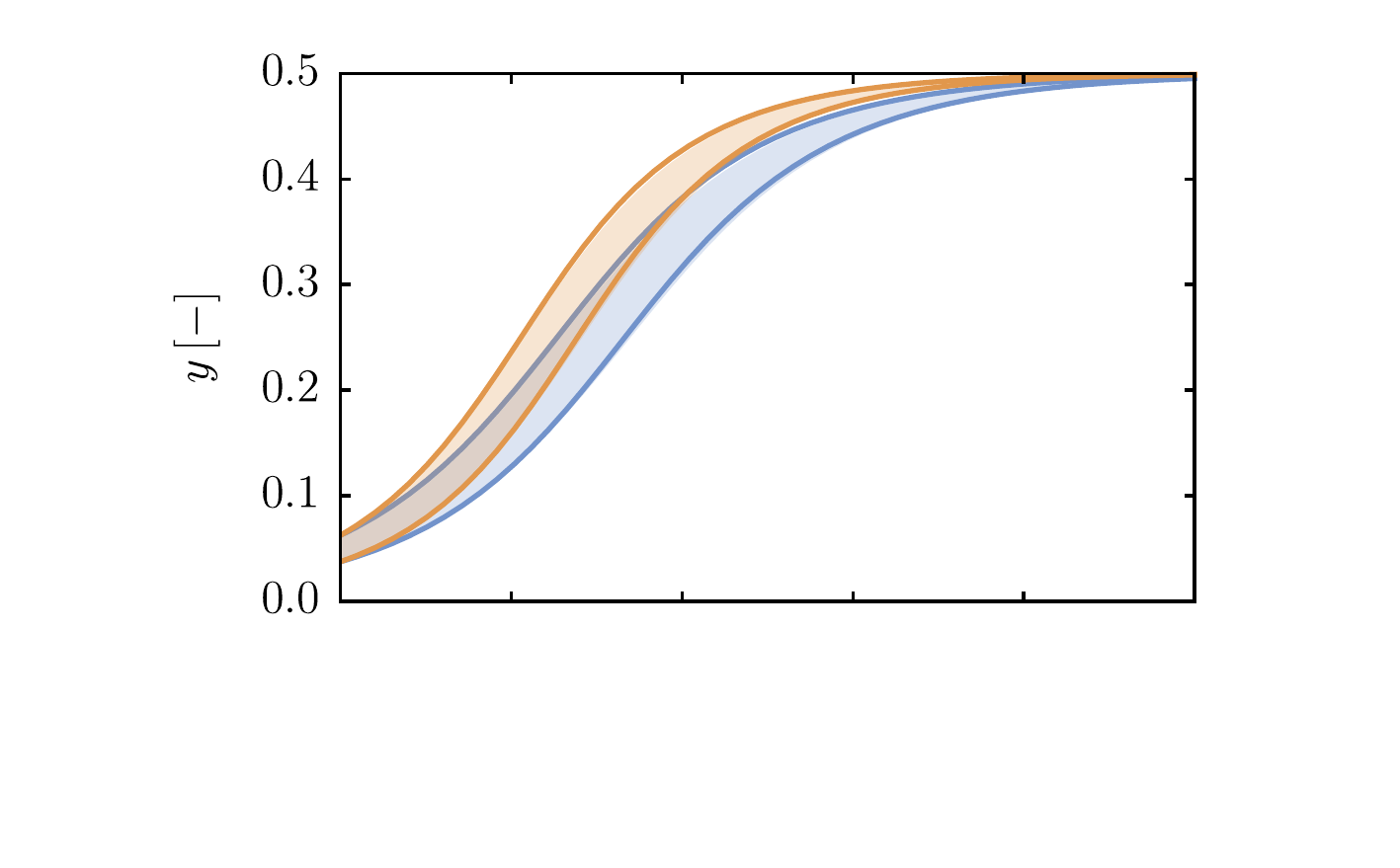}}
\put(187,396){(A)}
\put(119,131){$\mathrm{max\,SE}$}
\put(134,91){$\mathrm{max\,SR}$}
\put(187,266){(B)}
\put(187,136){(C)}
\end{picture}
%}
\caption{(Color Online) (A)~Bandwidth areas (colored areas), dispersion bands (lines), (B)~ideal separation efficiency (red), resolution index (black, dashed is \mbox{$2\times\mathrm{RI}$}), and (C)~separation efficiency (black), separation resolution (red) as function of time $t$.
The initial conditions are \mbox{$y_0^{}=0.05$}, \mbox{$m_0^r=1$}, \mbox{$s_0^{rr}=0$}, \mbox{$s_0^{yr}=0$}, and \mbox{$\sqrt{s^{yy}_0}=\mathrm{BW}(0)=50\%\,y_0^{}$}.
The values for $\alpha$ are \mbox{$\alpha_1^{}=1$} (blue) and \mbox{$\alpha_2^{}=1.2$} (orange).
}
\label{fig:indicators}
\end{figure}
Figure~\ref{fig:indicators} shows the results in terms of the Yang's indicators~\cite{yang2018} and those above defined.
Panel~A shows the superposition of the two ``dispersed'' particle streams and the comparison between the bandwidth areas and the dispersion bands, which results very close each other as in the limit of small perturbations.
Panel~B shows Yang's ideal separation efficiency $\mathrm{ISE}$ and the resolution index $\mathrm{RI}$, which have very distinct behaviors.
Finally, panel~C show the comparison of Yang's separation efficiency $\mathrm{SE}$ and the separation resolution $\mathrm{SR}$.
Also for these two indicators the behaviors are different. It is even more important to notice that the maximum in $\mathrm{SE}$ and $\mathrm{SR}$ do not coincide. Specifically, the separation resolution anticipates the position for the optimal ``section'' where the two particle streams are better separated (see Sec.~\ref{sec:discussion} for a discussion).
Therefore, for that regards the separation indicators the two methods give different results.

\section{Discussion}\label{sec:discussion}
The method used by Yang and Vitali is conceptually similar, but not identical, to the method proposed by Garofalo. They give practically identical results in terms of bandwidth and dispersion band in the case of univariate (single parameter) and small perturbation analysis.
However, when the Yang's approach is applied to multiparametric sensitivity analysis, such as illustrated in Fig.~\ref{fig:comparison3}, and when the indicators for the two methods are compared, the lack of statistical dispersion analysis gives different results.

\begin{figure}[!!t]
%\fbox{
\begin{picture}(226,155)
\put(-40,-15){\includegraphics[width=10cm]{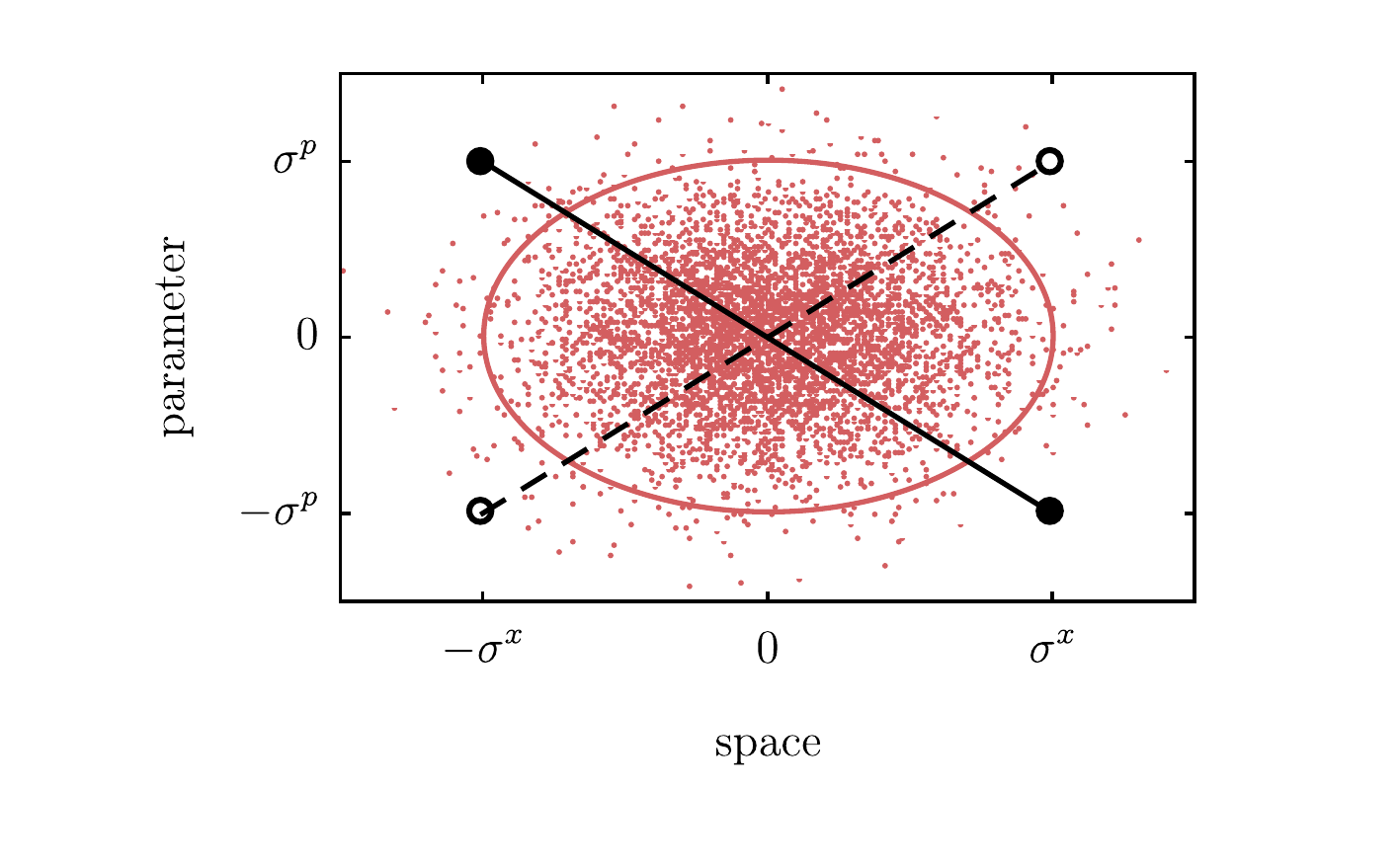}}
%\put(187,271){(A)}
%\put(187,136){(B)}
\end{picture}
%}
\caption{(Color Online) Example of sample statistics for randomly picked particles with spatial dispersion $\sigma^x_{}$ and parameter dispersion $\sigma^p_{}$ (dots).
The red ellipse is associated with the covariance matrix in Garofalo's method, the dots are associated with the Yang's method and they represent the extremal values.
}
\label{fig:explanation}
\end{figure}
In the Yang's method a distributed population with mean $\bm{m}$ and covariance matrix $\bm{s}$ is represented just by two extremal values, and for that Yang's method assumes a different statistics for the sample.
Figure~\ref{fig:explanation} shows the representations of this fact in a simplified way. A particle population (dots) distributed in both space and parameter is represented by its mean and variance in statistics-based method (ellipse), while it is represented by extremal values (empty/filled circles) in the case of the kinematic-based method by Yang et al. Note also that in the case of Yang's method it must be established a priori what is the combination for the ``worst'' and ``best'' cases, while in statistics-based method this is not necessary.
It is even more important to recognize that in statistical dispersion analisys the parameter dispersion $\bm{s^{pp}_0}$ is involved in the dynamics of spatial dispersion $\bm{s^{xx}_{}}$ by means of the cross-covariance $\bm{s^{xp}_{}}$ one-way coupling.

\begin{figure}[!!t]
%\fbox{
\begin{picture}(226,155)
\put(-40,-15){\includegraphics[width=10cm]{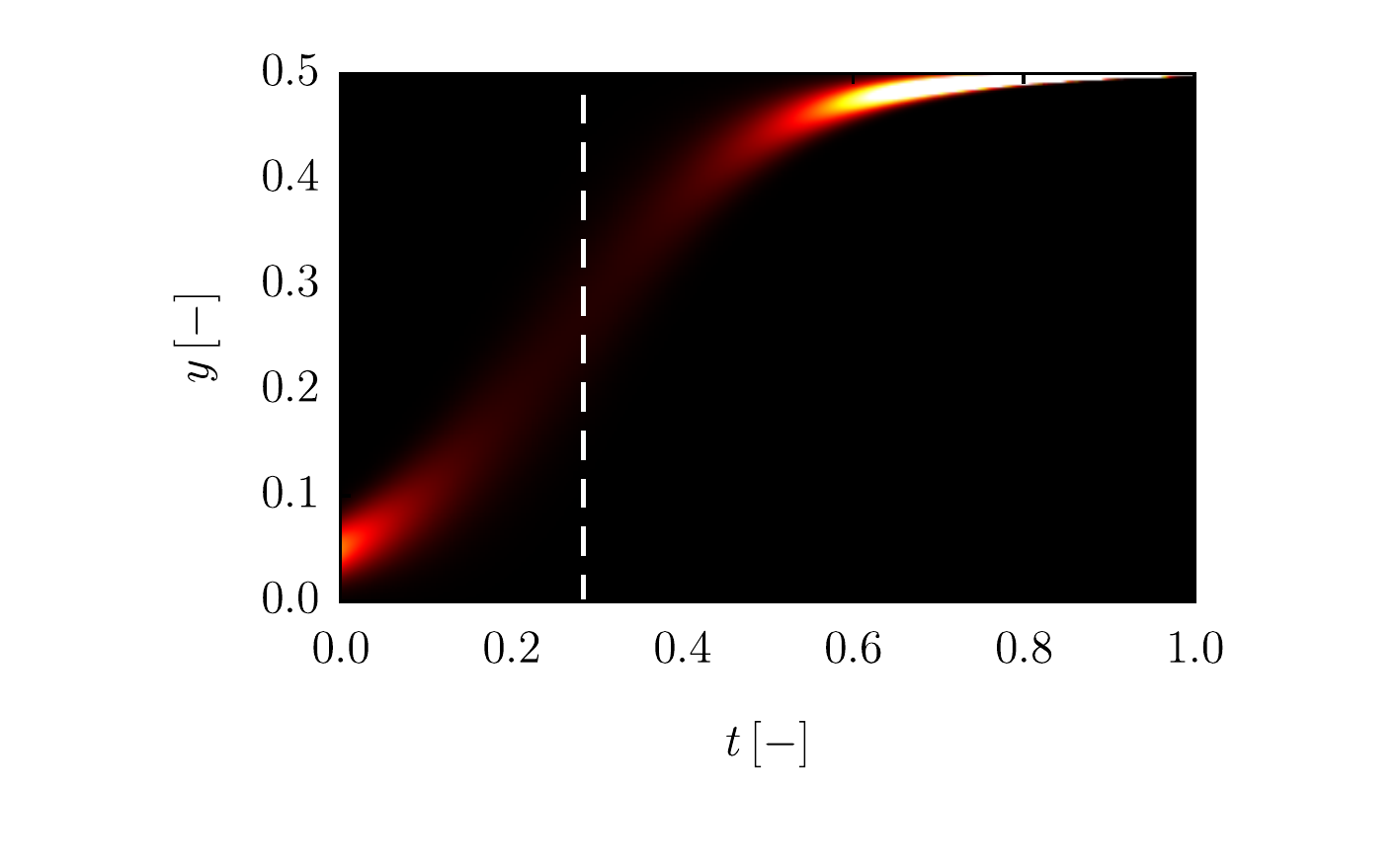}}
%\put(187,271){(A)}
%\put(187,136){(B)}
\end{picture}
%}
\caption{(Color Online) Product of the spatial marginals for the case reported in Fig.~\ref{fig:indicators} as function of the time $t$ and the cross-sectional position $y$.
The vertical line correspond to the $\mathrm{max\,SR}$ line depicted in Fig.~\ref{fig:indicators}(C).
}
\label{fig:pdfsproduct}
\end{figure}

Another issue with the Yang's method is that it does not provide information on the particle number density.
The discrepancies observed between the information provided by $\mathrm{SE}$ and $\mathrm{SR}$ are a consequence of this lack.
Indeed, in the statistical method the product in the integral Eq.~\eqref{eq:sr} gives a measure of the superposition of the two particle streams in terms of position, dispersion and particle number density of the two streams, see Fig.~\ref{fig:pdfsproduct}.
Conversely, the Yang's method takes into account just for the position and the dispersion of the two particle streams.
It would be possible to include the particle number density by randomly picking particles in the coordinate+parameter space and evolve the corresponding trajectories~\cite{gergely2017}.
However, as discussed in~\cite{garofalo2018} this approach is computationally inconvenient, and the Garofalo's method supersedes admirably this type of simulations. Furthermore, the generalization of Yang's indicators is not straigthforward in the case of distributed populations.

Finally, the information provided by the particle number density is essential when the characterization method aims (i)~to quantify the separation efficiency (as defined in~\cite{garofalo2018}), (ii)~to measure single-value property, or (iii)~to infer parameter histograms.
Applications of the statistic-based method for acoustophoresis separations have been thoroughly investigated for arbitrary statistics in~\cite{garofalo2018}.

%\put(187,136){\begin{color}{white}(D)\end{color}}

% As the particle swarm reaches the outlet section at \mbox{$x=L$} for \mbox{$t^{}_\mathrm{sep}=t^{}_\mathrm{sep}(L)$}, the fraction of the particles entering the side outlet, or side-stream recovery, can be computed by considering the $y$-marginal,
% \begin{equation}
% \rho^y_{}(y,t)=\mathcal{N}[y\,|\,m^y_{}(t),s^{yy}_{}(t)]\,,
% \end{equation}
% and its cumulative density function (CDF)
% \begin{equation}\label{eq:pdf}
% \mathrm{SSR}=\frac{1}{2}\mathrm{erfc}\left[\frac{m^y_{}(t^{}_\mathrm{sep})-y_\mathrm{sep}^{}}{\sqrt{2\,s^{yy}_{}(t_\mathrm{sep}^{})}}\right]\,,
% \end{equation}
% where $m^y_{}$ is the $y$-component of $\bm{m}$, $y_\mathrm{sep}^{}$ is the position where the separation between the center and side outlet occurs, and $s^{yy}_{}$ is the dispersion (variance) along the $y$-direction.

% \paragraph*{Multiparametric Analysis.}~Albeit feasible, these two approaches to the multiparametric sensitivity analysis of the trajectories are cumbersome when the number of parameters becomes large.

\section{Concluding Remarks}
In this communication we harmonize the approach used by Yand and Vitali for the study of the parameters influence in acoustophoretic separation with the statistic-based method in~\cite{garofalo2014,garofalo2018}.
We recognize a partial equivalence of the two method by (i)~connecting the mean-and-covariance dynamics with the perturbative approach adopted by Yang et al. for the case of small perturbation and (ii)~comparing the heuristic indicators they introduced with indicators derived by adopting the statistic approach.
Furthermore, we addressed the limitations of a statistic-less method when applied to the characterization of processes in which distributed parameters are involved.

Ultimately, it can resume that the methodology introduced by Yang et al. coincides with the method proposed by Garofalo when statistic-less and univariate small perturbations are considered.
However, the Yang's method does not extend straightforwardly in the case of multivariate parametric sensitivity analysis, and when the information of the particle number density is requested.
For these reasons it claims that the method proposed by Garofalo's method is more general and has wider application opportunities.

\end{document}